\documentclass[aps,twocolumn,floatfix,showpacs,a4paper,twoside]{revtex4}
\usepackage{graphicx}

\begin{document}

\title{Nanolithography with metastable helium atoms in a high-power\\standing-wave light field}

\author{S.J.H.~Petra}
\author{L.~Feenstra}\thanks{\emph{Current address}: Physikalisches Institut, Universit\"at Hei\-del\-berg, Philosophenweg 12, 69120 Heidelberg, Germany.}
\author{W.~Hogervorst}
\author{W.~Vassen}

\affiliation{Atomic and Laser Physics Group, Laser Centre Vrije Universiteit, De Boelelaan 1081, 1081 HV Amsterdam, The Netherlands\\(Fax: +31-20/44-47999, E-mail: stefan@nat.vu.nl)}

\begin{abstract}
We have created periodic nanoscale structures in a gold substrate with a lithography process using metastable triplet helium atoms that damage a hydrofobic resist layer on top of the substrate.
A beam of metastable helium atoms is transversely cooled and guided through an intense standing-wave light field.
Compared to commonly used low-power optical masks, a high-power light field (saturation parameter of $10^7$) increases the confinement of the atoms in the standing-wave considerably, and makes the alignment of the experimental setup less critical.
Due to the high internal energy of the metastable helium atoms (20~eV), a dose of only one atom per resist molecule is required.
With an exposure time of only eight minutes, parallel lines with a separation of 542~nm and a width of 100~nm (1/11th of the wavelength used for the optical mask) are created.
\end{abstract}


\maketitle

\section{Introduction}
\label{intro}
Since the beginning of the previous decade, atom-optical manipulation of neutral atoms has been used to create nanoscale structures (for a recent review of atom nanofabrication, see Meschede and Metcalf \cite{mesc03}).
Two different processes can be distinguished for the fabrication of such structures.
In a deposition process nanoscale structures are directly grown on a surface.
Direct deposition experiments have been performed with sodium \cite{timp92}, chromium \cite{mccl93} and aluminum \cite{mcgo95}.
In a lithography procedure, on the other hand, the atoms first locally damage an organic resist layer and next the pattern is transferred to an underlying metal film in a wet-etching process \cite{xia95}.
Although this last technique is a two-step process, it has the advantage that in principle background-free structures can be created, and feature broadening caused by the growth process is avoided.
To create a pattern on the sample, a mechanical mask (e.g.\ an electron microscope grid) or an optical mask (standing light wave) can be used.
A mechanical mask partly blocks the atomic beam, and the grid structure is imaged onto the sample.
Using grids in a lithography process, microscale structures have been created with metastable argon \cite{berg95} and metastable helium \cite{nowa96,lu98}, showing an edge resolution of 100~nm and 40~nm respectively.
Atom lithography with an optical mask has been demonstrated with cesium \cite{liso97}, metastable argon \cite{john98}, and metastable neon \cite{enge99}.
In the cesium and argon experiments, structures with a feature size of respectively 120~nm and 65~nm were realized.
In the latter experiment the atoms were optically quenched by the light field and an exposure time of eight hours was required.   
In this paper we report lithography measurements with a mechanical and an optical mask for metastable helium atoms.
Helium has the highest internal energy (20~eV) of all metastable atoms.
Therefore, the dose of atoms required to damage a resist molecule can be small, compared to other metastable species.
This benefits the exposure time of the sample to the atomic beam.

An optical mask, formed by a standing-wave light field, acts as an array of cylindrical lenses, attracting the atoms to the potential minima of the light field.
Two different regimes can be distinguished.
A low-power regime, where the atoms are focused at the center of the Gaussian light beam, and a high-power regime, where the atoms oscillate around the potential minima of the light field.
The lithography experiments mentioned above were all performed in the focusing regime using low-power optical masks.
Theoretically, this regime can be described straightforwardly, since an analytical expression for the dipole force can be derived from the optical potential of the light field \cite{ashk78}.

We have performed lithography experiments with a high-power standing-wave light field, where the simple expression for the dipole force is not valid.
This is due to the fact that the dipole force becomes velocity dependent, because the transverse velocity of the atoms, channeling through the standing-wave light field, can no longer be neglected.
In this high-power regime, the dipole force can be described by a numerical model, first presented by Minogin and Serimaa \cite{mino79}.
In a previous paper we have compared this model with the model that describes the conventional dipole force \cite{petr03}.
In the high-power regime, the background due to non-focused atoms is significantly reduced compared to the focusing regime.
Also the alignment of the standing-wave laser beam with respect to the sample is far less critical.
The sample can be placed behind the center of the laser beam, which reduces disturbances of the standing wave pattern due to diffraction of the laser beam on the sample edge significantly \cite{ande99}.
Numerically calculated trajectories of atoms channeling through a standing-wave light field show an increased confinement of the atoms in the optical well along the atomic beam axis, due to a dissipative force that reduces the transverse velocity of the atoms.
The Full Width at Half Maximum (FWHM) of the atom distribution on the sample is calculated to be 40~nm for parameters as used in our experiment \cite{petr03}.

\section{Experimental setup}
\label{expsetup}

\begin{figure}[t]
\includegraphics[width=\columnwidth,keepaspectratio]{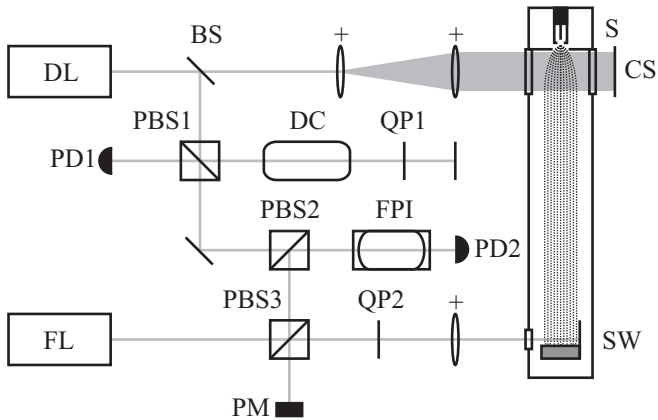}
\caption
{\label{setup}
Schematic of the setup.
Metastable helium atoms are produced in DC discharge source S\@.
Diode laser DL is used for transverse cooling of the atomic beam in collimation section CS\@.
Part of the diode laser output (4\%) is split by beam splitter BS for spectroscopy.
After reflection by polarizing beam splitter PBS1 and double-passing discharge cell DC and quarter-wave plate QP1, the light is detected by photodiode PD1\@.
The light transmitted by PBS1 also passes PBS2 and is analyzed with Fabry-Perot interferometer FPI and photodiode PD2\@.
The beam from fiber laser FL is partly reflected by PBS3 and also analyzed with FPI\@.
The light transmitted by PBS3 is retro-reflected by mirror SW, that is mounted on the sample holder inside the vacuum chamber.
After double-passing QP2, the light of the standing wave is reflected by PBS3 and monitored with power meter PM\@.}
\end{figure}

The experimental setup is shown in figure~\ref{setup}.
The metastable helium atoms are produced in a water-cooled DC discharge \cite{rooi96}.
In the discharge helium atoms are transferred from the $1~^1\textrm{S}_0$ ground state to the $2~^3\textrm{S}_1$ metastable state, which has a lifetime of 8000~s.
The atoms have a mean longitudinal velocity of 2000~m/s, which corresponds to an energy of 0.08~eV.
The atomic beam is transversely cooled in two dimensions using curved-wavefront Doppler cooling \cite{rooi96,aspe90}.
The collimation of the atomic beam reduces the transverse velocity spread of the atoms to about 3~m/s, corresponding to a beam divergence of 9~mrad.
In the high-power regime, the guiding of the atoms through the standing wave is not very sensitive to the transverse velocity of the atoms, and calculations have shown that the divergence is sufficiently small \cite{petr03}.
The collimation of the beam enhances the flux of the atomic beam by a factor of 6 to $1.1 \times 10^{10}$ atoms\,s$^{-1}$\,mm$^{-2}$.
The light for transverse cooling (20~mW) is provided by a DBR (Distributed Bragg Reflector) diode laser.
This laser is locked to the $2~^3\textrm{S}_1 \rightarrow 2~^3\textrm{P}_2$ transition of the helium atom using Doppler-free saturation spectroscopy with a lock-in technique.
The wavelength of the light corresponding to this transition is 1083~nm.
The bandwidth of the laser is experimentally determined to be 2~MHz, which is comparable to the 1.6~MHz linewidth of the $2~^3\textrm{P}_2$ excited state. 

The standing-wave light field is obtained by retro-re\-flect\-ing an 800~mW light beam, provided by a fiber laser (IPG, model YLD-1BC).
This laser also operates at a wavelength of 1083~nm and is 375~MHz blue detuned from the $2~^3\textrm{S}_1 \rightarrow 2~^3\textrm{P}_2$ transition to reduce spontaneous emission of the atoms in the standing-wave light field.
Higher detunings are unfavorable, because the atoms can then also interact with the light via the $2~^3\textrm{S}_1 \rightarrow 2~^3\textrm{P}_1$ transition.
The frequency of this transition is only 2.3~GHz higher than the frequency of the $2~^3\textrm{S}_1 \rightarrow 2~^3\textrm{P}_2$ transition.

The fiber laser is unlocked but its frequency with respect to the locked diode laser is monitored with a Fabry-Perot interferometer, whose cavity length is scanned with a piezoelectric transducer.
The fiber laser has a measured short-term linewidth of 1~MHz and a good long term stability.
Only drifts on the order of 10~MHz are visible with respect to the locked diode laser during the exposure times.
The laser beam is focused onto a retro-reflecting mirror to a size ($1/\textrm{e}^2$ radius) of 0.33~mm, which translates to a saturation parameter (defined as the intensity of the light divided by the saturation intensity of the atomic transition) of $10^7$.
The retro-reflecting mirror and the sample are mechanically fixed in their relative position onto the sample holder, which is made of a solid aluminum block.
For each experiment a substrate is mounted on the sample holder in free air.
The sample holder is then introduced into the vacuum chamber via an air-lock.
The exposure dose is determined by measuring the current through a fixed Faraday cup due to the metastable atoms, before and after the exposure \cite{rooi96}.

The laser beam is clipped by the sample, located $100~\mu$m from the center of the laser beam, downstream the atomic beam.
Numerical calculations that take into account the longitudinal and transverse velocity spread of the atomic beam have shown that this is the best position for the sample \cite{petr03}.

For proper alignment of the standing wave, the laser beam is first adjusted such that it is not clipped by the sample and thus 100\% of the light is back-reflected by the standing-wave mirror onto the power meter (see figure~\ref{setup}). 
Next, the beam is moved by a micrometer translation stage (not indicated in figure~\ref{setup}) and clipped by the sample such that only 50\% of the light is back-reflected, which indicates that the sample position is at the center of the laser beam.
Finally, additional fine tuning is done with the translation stage to move the laser beam to the desired position with respect to the sample.

The samples used in the experiment are silicon substrates covered with a 1~nm chromium layer and a 30~nm gold layer.
These samples are cleaned with pure ethanol and with a solution of concentrated sulfuric acid (98\% pure) and hydrogen peroxide (30\% pure).
In the experiments two different resist layers are used: nonanethiol [CH$_3$(CH$_2$)$_8$SH] and dodecanethiol [CH$_3$(CH$_2$)$_{11}$SH].
A Self-Assembled Monolayer (SAM) of either of these alkanethiols is deposited onto the substrates by leaving the samples overnight in a 1~mM solution of the alkanethiol in ethanol.
The SAM forms a hydrofobic resist layer that protects the gold layer in the wet etching process.

For nonanethiol and dodecanethiol the samples are exposed to the atomic beam for eight minutes and ten minutes respectively.
This corresponds to a dose of about one metastable atom per resist molecule.
During the exposure, the SAM is locally damaged and loses its hydrofobic character.
After being exposed, the samples are wet etched in a cyanide etching solution \cite{nowa96}.
The etching times for nonanethiol and dodecanethiol are also eight and ten minutes respectively.
In the etching process, the SAM molecules that are damaged due to the impact of a metastable helium atom are removed, as well as the gold film underneath these damaged molecules.
Longer etching times can affect areas untouched by the metastable helium atoms \cite{xia95}.
The samples are finally analyzed with an Atomic Force Microscope (AFM).

\section{Results and discussion}
\label{results}

\begin{figure}[t]
\includegraphics[width=\columnwidth,keepaspectratio]{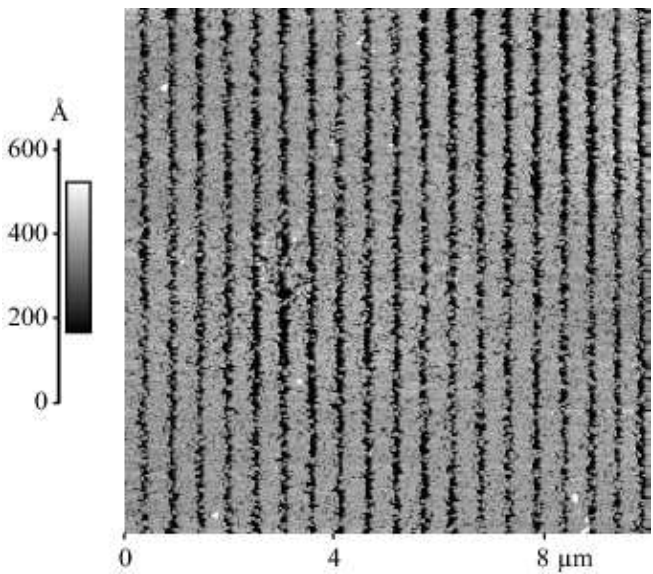}
\caption
{\label{lithoplaatje}
AFM image of a $10~\mu\textrm{m} \times 10~\mu\textrm{m}$ scan.
The dark lines indicate the regions where the metastable helium atoms have hit the sample.
The separation between two successive lines is 542~nm, which is half of the wavelength used for the standing wave.}
\end{figure}

\begin{figure}[t]
\includegraphics[width=\columnwidth,keepaspectratio]{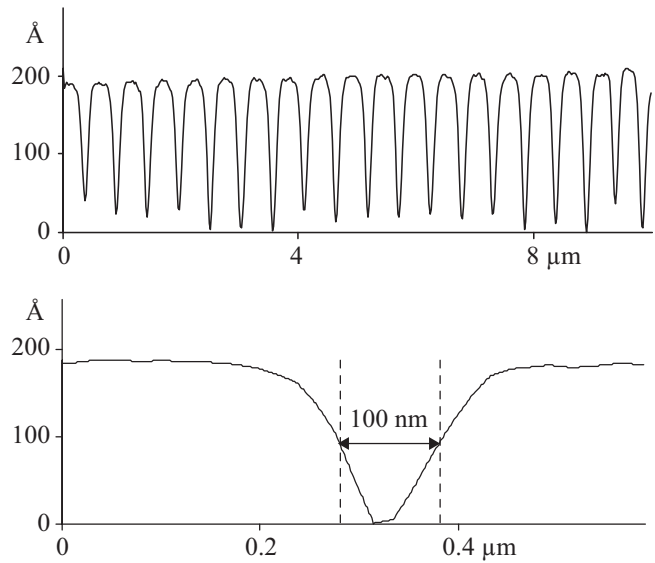}
\caption
{\label{heightprofile}
Height profiles of the AFM image of figure~\ref{lithoplaatje}.
The upper figure shows the average height profile of the entire image.
The lower figure shows that the average FWHM of a single line is 100~nm.}
\end{figure}

A typical result of a sample image produced with a standing-wave light field is shown in figure~\ref{lithoplaatje}.
The image shows a $10~\mu\textrm{m} \times 10~\mu\textrm{m}$ scan of the substrate, taken with an AFM\@.
The black lines indicate the positions where the metastable helium atoms have hit the sample and where the gold layer is removed in the etching process.
The average distance between two successive lines is 542~nm, which is exactly half of the wavelength of the standing-wave light field (1083~nm).
A more quantitative analysis of the pattern is made in figure~\ref{heightprofile}.
This figure shows that the average FWHM of the lines in figure~\ref{lithoplaatje} is about 100~nm.
The results are the same for both the nonanethiol and dodecanethiol SAM, although the latter requires a 25\% longer exposure and etching time.
The irregular structure of the black lines is probably caused by the graininess of the gold layer, which is also visible on AFM images of untreated samples.
For samples that are under-exposed and/or under-etched, the black lines are broken, while samples that are over-exposed and/or over-etched show connections between two successive lines.
The chemistry involved in the experiments makes the reproducibility of the measurements difficult.
In some cases no line structures were found on the sample although all physical parameters of the experiment were kept identical.
When a sample is visibly inspected by eye, diffracted light indicates that the sample contains well written nanostructures.

\begin{figure}[t]
\includegraphics[width=\columnwidth,keepaspectratio]{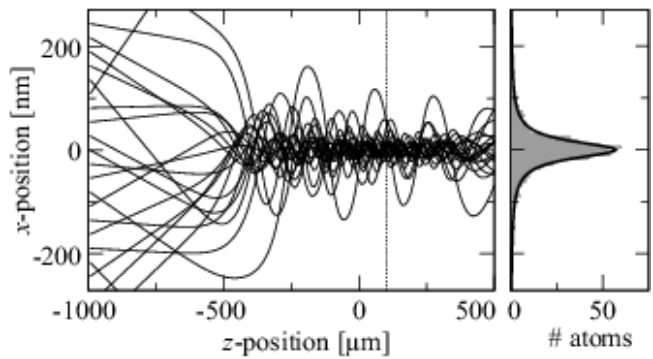}
\caption
{\label{trajectories}
Results of numerically calculated trajectories of atoms traveling through the standing-wave light field \cite{petr03}.
The left-hand graph shows simulated trajectories of metastable helium atoms with a mean longitudinal velocity of 2000~m/s and a transverse velocity spread of 3~m/s channeling through a standing-wave light field with a power of 800~mW and a detuning of 375~MHz.
The dashed line indicates the sample position at $100~\mu$m from the center of the laser beam, downstream the atomic beam.
The right-hand graph shows a histogram of the atom distribution on the sample.
The FWHM of the distribution taken at the sample position is 40~nm.}
\end{figure}

\begin{figure}[t]
\includegraphics[width=\columnwidth,keepaspectratio]{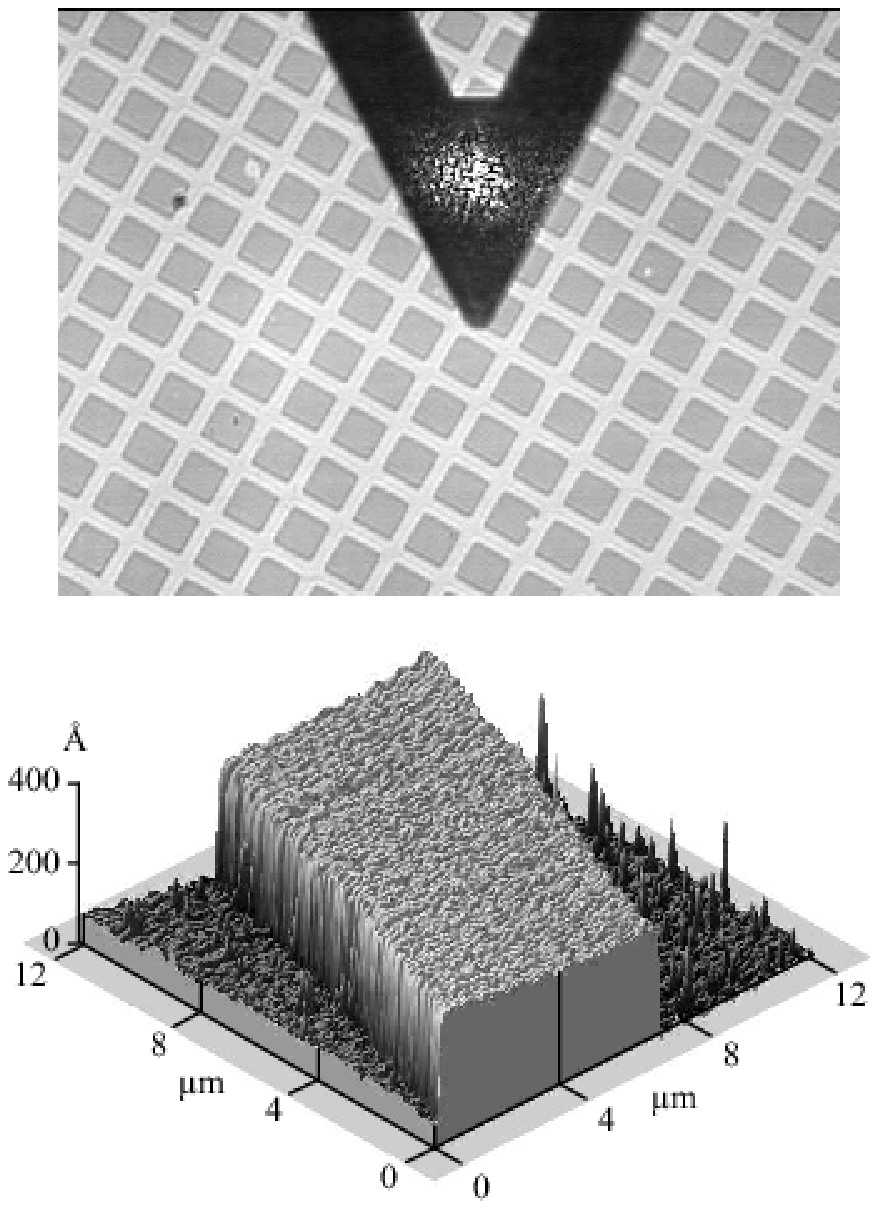}
\caption
{\label{afmgrid}
Results of atom lithography with a 1000~lines/inch mechanical mask.
The upper figure shows an optical microscope image (top view) of a sample and the AFM tip (pointing into the paper).
The lower image is an AFM scan of one bar of the grid structure.
Note the scale difference between the horizontal and vertical axes.
The side wall inclination turns out to be about $40^{\circ}$, corresponding to a step-edge of 40~nm.}
\end{figure}

To compare the measured structures with theory, numerical simulations using the Minogin and Serimaa model are shown in figure~\ref{trajectories}.
The input parameters of these calculations correspond to the values used in the experiment.
Based on these calculations, a FWHM of a single line of 40~nm is expected, which is a factor of 2.5 smaller than observed experimentally.

The discrepancy between the measured and the calculated width is attributed to both the graininess of the gold layer and the wet-etching process, which is an isotropic process that also removes the gold layer from inside a line structure during etching.
This conclusion is in agreement with experiments performed with a mechanical mask, shown in figure~\ref{afmgrid}.
In these measurements a grid of 1000~mesh (1000~lines per inch) is used instead of an optical mask.
From analysis with the AFM a $40^{\circ}$ inclination in the gold layer is found, in agreement with an earlier observation \cite{nowa96}.
The AFM tip has a cone angle of $10^{\circ}$ and does therefore not limit the measured angle.

The $40^{\circ}$ side wall inclination translates to a step-edge of 40~nm and increases the expected FWHM linewidth of a single line with 40~nm.
Nanolithography with a wet-etching technique incorporating a SAM as a resist layer therefore seems to be limited to structure sizes larger than 50~nm.

Comparing metastable helium nanolithography in the \linebreak high-power optical mask regime with results employing other elements shows some clear advantages.
First, the high internal energy of metastable helium atoms allows a relatively short exposure time of eight minutes.
Second, no pedestal due to non-focused atoms is visible.
Third, the high power available nowadays from fiber lasers allows an extremely large saturation parameter ($10^7$) and deep penetration in the channeling regime.
In this regime a feature size of 1/11th of the optical-mask wavelength is observed, which is hard to realize in the focusing (low-power) regime, due to a large sensitivity to the longitudinal and transversal velocity spread in the atomic beam in this regime, and because the sample position with respect to the standing wave is very critical.

\begin{acknowledgments}
The authors thank K.A.H.~van~Leeuwen for fruitful discussions.
Financial support from the Foundation for Fundamental Research on Matter (FOM) is gratefully acknowledged.
\end{acknowledgments}


\begin{thebibliography}{17}
\expandafter\ifx\csname natexlab\endcsname\relax\def\natexlab#1{#1}\fi
\expandafter\ifx\csname bibnamefont\endcsname\relax
  \def\bibnamefont#1{#1}\fi
\expandafter\ifx\csname bibfnamefont\endcsname\relax
  \def\bibfnamefont#1{#1}\fi
\expandafter\ifx\csname citenamefont\endcsname\relax
  \def\citenamefont#1{#1}\fi
\expandafter\ifx\csname url\endcsname\relax
  \def\url#1{\texttt{#1}}\fi
\expandafter\ifx\csname urlprefix\endcsname\relax\def\urlprefix{URL }\fi
\providecommand{\bibinfo}[2]{#2}
\providecommand{\eprint}[2][]{\url{#2}}

\bibitem[{\citenamefont{Meschede and Metcalf}(2003)}]{mesc03}
\bibinfo{author}{\bibfnamefont{D.}~\bibnamefont{Meschede}} \bibnamefont{and}
  \bibinfo{author}{\bibfnamefont{H.}~\bibnamefont{Metcalf}},
  \bibinfo{journal}{J. Phys. D} \textbf{\bibinfo{volume}{36}},
  \bibinfo{pages}{R17} (\bibinfo{year}{2003}).

\bibitem[{\citenamefont{Timp et~al.}(1992)\citenamefont{Timp, Behringer,
  Tennant, Cunningham, Prentiss, and Berggren}}]{timp92}
\bibinfo{author}{\bibfnamefont{G.}~\bibnamefont{Timp}},
  \bibinfo{author}{\bibfnamefont{R.~E.} \bibnamefont{Behringer}},
  \bibinfo{author}{\bibfnamefont{D.~M.} \bibnamefont{Tennant}},
  \bibinfo{author}{\bibfnamefont{J.~E.} \bibnamefont{Cunningham}},
  \bibinfo{author}{\bibfnamefont{M.}~\bibnamefont{Prentiss}}, \bibnamefont{and}
  \bibinfo{author}{\bibfnamefont{K.~K.} \bibnamefont{Berggren}},
  \bibinfo{journal}{Phys. Rev. Lett.} \textbf{\bibinfo{volume}{69}},
  \bibinfo{pages}{1636} (\bibinfo{year}{1992}).

\bibitem[{\citenamefont{McClelland et~al.}(1993)\citenamefont{McClelland,
  Scholten, Palm, and Celotta}}]{mccl93}
\bibinfo{author}{\bibfnamefont{J.~J.} \bibnamefont{McClelland}},
  \bibinfo{author}{\bibfnamefont{R.~E.} \bibnamefont{Scholten}},
  \bibinfo{author}{\bibfnamefont{E.~C.} \bibnamefont{Palm}}, \bibnamefont{and}
  \bibinfo{author}{\bibfnamefont{R.~J.} \bibnamefont{Celotta}},
  \bibinfo{journal}{Science} \textbf{\bibinfo{volume}{262}},
  \bibinfo{pages}{877} (\bibinfo{year}{1993}).

\bibitem[{\citenamefont{McGowan et~al.}(1995)\citenamefont{McGowan, Giltner,
  and Lee}}]{mcgo95}
\bibinfo{author}{\bibfnamefont{R.~W.} \bibnamefont{McGowan}},
  \bibinfo{author}{\bibfnamefont{D.~M.} \bibnamefont{Giltner}},
  \bibnamefont{and} \bibinfo{author}{\bibfnamefont{S.~A.} \bibnamefont{Lee}},
  \bibinfo{journal}{Opt. Lett.} \textbf{\bibinfo{volume}{20}},
  \bibinfo{pages}{2535} (\bibinfo{year}{1995}).

\bibitem[{\citenamefont{Xia et~al.}(1995)\citenamefont{Xia, Zhao, Kim, and
  Whitesides}}]{xia95}
\bibinfo{author}{\bibfnamefont{Y.}~\bibnamefont{Xia}},
  \bibinfo{author}{\bibfnamefont{X.-M.} \bibnamefont{Zhao}},
  \bibinfo{author}{\bibfnamefont{E.}~\bibnamefont{Kim}}, \bibnamefont{and}
  \bibinfo{author}{\bibfnamefont{G.~M.} \bibnamefont{Whitesides}},
  \bibinfo{journal}{Chem. Mater.} \textbf{\bibinfo{volume}{7}},
  \bibinfo{pages}{2332} (\bibinfo{year}{1995}).

\bibitem[{\citenamefont{Berggren et~al.}(1995)\citenamefont{Berggren, Bard,
  Wilbur, Gillaspy, Helg, McClelland, Rolston, Phillips, Prentiss, and
  Whitesides}}]{berg95}
\bibinfo{author}{\bibfnamefont{K.~K.} \bibnamefont{Berggren}},
  \bibinfo{author}{\bibfnamefont{A.}~\bibnamefont{Bard}},
  \bibinfo{author}{\bibfnamefont{J.~L.} \bibnamefont{Wilbur}},
  \bibinfo{author}{\bibfnamefont{J.~D.} \bibnamefont{Gillaspy}},
  \bibinfo{author}{\bibfnamefont{A.~G.} \bibnamefont{Helg}},
  \bibinfo{author}{\bibfnamefont{J.~J.} \bibnamefont{McClelland}},
  \bibinfo{author}{\bibfnamefont{S.~L.} \bibnamefont{Rolston}},
  \bibinfo{author}{\bibfnamefont{W.~D.} \bibnamefont{Phillips}},
  \bibinfo{author}{\bibfnamefont{M.}~\bibnamefont{Prentiss}}, \bibnamefont{and}
  \bibinfo{author}{\bibfnamefont{G.~M.} \bibnamefont{Whitesides}},
  \bibinfo{journal}{Science} \textbf{\bibinfo{volume}{269}},
  \bibinfo{pages}{1255} (\bibinfo{year}{1995}).

\bibitem[{\citenamefont{Nowak et~al.}(1996)\citenamefont{Nowak, Pfau, and
  Mlynek}}]{nowa96}
\bibinfo{author}{\bibfnamefont{S.}~\bibnamefont{Nowak}},
  \bibinfo{author}{\bibfnamefont{T.}~\bibnamefont{Pfau}}, \bibnamefont{and}
  \bibinfo{author}{\bibfnamefont{J.}~\bibnamefont{Mlynek}},
  \bibinfo{journal}{Appl. Phys. B} \textbf{\bibinfo{volume}{63}},
  \bibinfo{pages}{203} (\bibinfo{year}{1996}).

\bibitem[{\citenamefont{Lu et~al.}(1998)\citenamefont{Lu, Baldwin, Hoogerland,
  Buckman, Senden, Sheridan, and Boswell}}]{lu98}
\bibinfo{author}{\bibfnamefont{W.}~\bibnamefont{Lu}},
  \bibinfo{author}{\bibfnamefont{K.~G.~H.} \bibnamefont{Baldwin}},
  \bibinfo{author}{\bibfnamefont{M.~D.} \bibnamefont{Hoogerland}},
  \bibinfo{author}{\bibfnamefont{S.~J.} \bibnamefont{Buckman}},
  \bibinfo{author}{\bibfnamefont{T.~J.} \bibnamefont{Senden}},
  \bibinfo{author}{\bibfnamefont{T.~E.} \bibnamefont{Sheridan}},
  \bibnamefont{and} \bibinfo{author}{\bibfnamefont{R.~W.}
  \bibnamefont{Boswell}}, \bibinfo{journal}{J. Vac. Sci. \& Tech.~B}
  \textbf{\bibinfo{volume}{16}}, \bibinfo{pages}{3846} (\bibinfo{year}{1998}).

\bibitem[{\citenamefont{Lison et~al.}(1997)\citenamefont{Lison, Adams,
  Haubrich, Kreis, Nowak, and Meschede}}]{liso97}
\bibinfo{author}{\bibfnamefont{F.}~\bibnamefont{Lison}},
  \bibinfo{author}{\bibfnamefont{H.-J.} \bibnamefont{Adams}},
  \bibinfo{author}{\bibfnamefont{D.}~\bibnamefont{Haubrich}},
  \bibinfo{author}{\bibfnamefont{M.}~\bibnamefont{Kreis}},
  \bibinfo{author}{\bibfnamefont{S.}~\bibnamefont{Nowak}}, \bibnamefont{and}
  \bibinfo{author}{\bibfnamefont{D.}~\bibnamefont{Meschede}},
  \bibinfo{journal}{Appl. Phys. B} \textbf{\bibinfo{volume}{65}},
  \bibinfo{pages}{419} (\bibinfo{year}{1997}).

\bibitem[{\citenamefont{Johnson et~al.}(1998)\citenamefont{Johnson, Thywissen,
  Dekker, Berg\-gren, Chu, Younkin, and Prentiss}}]{john98}
\bibinfo{author}{\bibfnamefont{K.~S.} \bibnamefont{Johnson}},
  \bibinfo{author}{\bibfnamefont{J.~H.} \bibnamefont{Thywissen}},
  \bibinfo{author}{\bibfnamefont{N.~H.} \bibnamefont{Dekker}},
  \bibinfo{author}{\bibfnamefont{K.~K.} \bibnamefont{Berg\-gren}},
  \bibinfo{author}{\bibfnamefont{A.~P.} \bibnamefont{Chu}},
  \bibinfo{author}{\bibfnamefont{R.}~\bibnamefont{Younkin}}, \bibnamefont{and}
  \bibinfo{author}{\bibfnamefont{M.}~\bibnamefont{Prentiss}},
  \bibinfo{journal}{Science} \textbf{\bibinfo{volume}{280}},
  \bibinfo{pages}{1583} (\bibinfo{year}{1998}).

\bibitem[{\citenamefont{Engels et~al.}(1999)\citenamefont{Engels, Salewski,
  Levsen, Sengstock, and Ertmer}}]{enge99}
\bibinfo{author}{\bibfnamefont{P.}~\bibnamefont{Engels}},
  \bibinfo{author}{\bibfnamefont{S.}~\bibnamefont{Salewski}},
  \bibinfo{author}{\bibfnamefont{H.}~\bibnamefont{Levsen}},
  \bibinfo{author}{\bibfnamefont{K.}~\bibnamefont{Sengstock}},
  \bibnamefont{and} \bibinfo{author}{\bibfnamefont{W.}~\bibnamefont{Ertmer}},
  \bibinfo{journal}{Appl. Phys. B} \textbf{\bibinfo{volume}{69}},
  \bibinfo{pages}{407} (\bibinfo{year}{1999}).

\bibitem[{\citenamefont{Ashkin}(1978)}]{ashk78}
\bibinfo{author}{\bibfnamefont{A.}~\bibnamefont{Ashkin}},
  \bibinfo{journal}{Phys. Rev. Lett.} \textbf{\bibinfo{volume}{40}},
  \bibinfo{pages}{729} (\bibinfo{year}{1978}).

\bibitem[{\citenamefont{Minogin and Serimaa}(1979)}]{mino79}
\bibinfo{author}{\bibfnamefont{V.~G.} \bibnamefont{Minogin}} \bibnamefont{and}
  \bibinfo{author}{\bibfnamefont{O.~T.} \bibnamefont{Serimaa}},
  \bibinfo{journal}{Opt. Commun.} \textbf{\bibinfo{volume}{30}},
  \bibinfo{pages}{373} (\bibinfo{year}{1979}).

\bibitem[{\citenamefont{Petra et~al.}(2003)\citenamefont{Petra, van Leeuwen,
  Feenstra, Hogervorst, and Vassen}}]{petr03}
\bibinfo{author}{\bibfnamefont{S.~J.~H.} \bibnamefont{Petra}},
  \bibinfo{author}{\bibfnamefont{K.~A.~H.} \bibnamefont{van Leeuwen}},
  \bibinfo{author}{\bibfnamefont{L.}~\bibnamefont{Feenstra}},
  \bibinfo{author}{\bibfnamefont{W.}~\bibnamefont{Hogervorst}},
  \bibnamefont{and} \bibinfo{author}{\bibfnamefont{W.}~\bibnamefont{Vassen}},
  \bibinfo{journal}{Eur. Phys. J.~D}  (\bibinfo{year}{2003}),
  \bibinfo{note}{online version 10.1140/epjd/e2003-00229-y}.

\bibitem[{\citenamefont{Anderson et~al.}(1999)\citenamefont{Anderson, Bradley,
  McClelland, and Celotta}}]{ande99}
\bibinfo{author}{\bibfnamefont{W.~R.} \bibnamefont{Anderson}},
  \bibinfo{author}{\bibfnamefont{C.~C.} \bibnamefont{Bradley}},
  \bibinfo{author}{\bibfnamefont{J.~J.} \bibnamefont{McClelland}},
  \bibnamefont{and} \bibinfo{author}{\bibfnamefont{R.~J.}
  \bibnamefont{Celotta}}, \bibinfo{journal}{Phys. Rev. A}
  \textbf{\bibinfo{volume}{59}}, \bibinfo{pages}{2476} (\bibinfo{year}{1999}).

\bibitem[{\citenamefont{Rooijakkers et~al.}(1996)\citenamefont{Rooijakkers,
  Hogervorst, and Vassen}}]{rooi96}
\bibinfo{author}{\bibfnamefont{W.}~\bibnamefont{Rooijakkers}},
  \bibinfo{author}{\bibfnamefont{W.}~\bibnamefont{Hogervorst}},
  \bibnamefont{and} \bibinfo{author}{\bibfnamefont{W.}~\bibnamefont{Vassen}},
  \bibinfo{journal}{Opt. Commun.} \textbf{\bibinfo{volume}{123}},
  \bibinfo{pages}{321} (\bibinfo{year}{1996}).

\bibitem[{\citenamefont{Aspect et~al.}(1990)\citenamefont{Aspect,
  Vansteenkiste, Kaiser, Haberland, and Karrais}}]{aspe90}
\bibinfo{author}{\bibfnamefont{A.}~\bibnamefont{Aspect}},
  \bibinfo{author}{\bibfnamefont{N.}~\bibnamefont{Vansteenkiste}},
  \bibinfo{author}{\bibfnamefont{R.}~\bibnamefont{Kaiser}},
  \bibinfo{author}{\bibfnamefont{H.}~\bibnamefont{Haberland}},
  \bibnamefont{and} \bibinfo{author}{\bibfnamefont{M.}~\bibnamefont{Karrais}},
  \bibinfo{journal}{Chem. Phys.} \textbf{\bibinfo{volume}{145}},
  \bibinfo{pages}{307} (\bibinfo{year}{1990}).

\end{thebibliography}
\end{document}